\documentclass[useAMS,usenatbib]{mn2e}

\usepackage{latexsym,graphicx,natbib}
\newcommand\kms{{\rm\,km\,s^{-1}}}

\newcommand\msun{\rm\,M_\odot}
\newcommand\rsun{\rm\,R_\odot}

\def\apgt{\ {\raise-.5ex\hbox{$\buildrel>\over\sim$}}\ }
\def\aplt{\ {\raise-.5ex\hbox{$\buildrel<\over\sim$}}\ }

\title[On the origin of high-velocity runaway stars]{On the origin of high-velocity runaway stars}
\author[V.V.Gvaramadze, A.Gualandris and S.Portegies Zwart]
       {Vasilii V. Gvaramadze$^{1}$\thanks{E-mail: vgvaram@mx.iki.rssi.ru},
       Alessia Gualandris$^{2}$\thanks{E-mail: alessiag@astro.rit.edu}
       and Simon Portegies Zwart$^{3}$\thanks{E-mail: spz@science.uva.nl}\\
       $^{1}$Sternberg Astronomical Institute, Moscow State University,
       Universitetskij Pr. 13, Moscow 119992, Russia\\
       $^{2}$Center for Computational Relativity and Gravitation,
       Rochester Institute of Technology, 85 Lomb Memorial Drive, Rochester NY 14623, USA\\
       $^{3}$Astronomical Institute `Anton Pannekoek' and Section
            Computational Science, Amsterdam University, Kruislaan 403,
            1098 SJ Amsterdam, the\\ Netherlands}
\begin{document}

\date{Accepted 2009 March 23. Received 2009 March 4}


\maketitle

\begin{abstract}
We explore the hypothesis that some high-velocity runaway stars
attain their peculiar velocities in the course of exchange
encounters between hard massive binaries and a very massive star
(either an ordinary $50-100 \, \msun$ star or a more massive one,
formed through runaway mergers of ordinary stars in the core of a
young massive star cluster). In this process, one of the binary
components becomes gravitationally bound to the very massive star,
while the second one is ejected, sometimes with a high speed. We
performed three-body scattering experiments and found that early
B-type stars (the progenitors of the majority of neutron stars) can
be ejected with velocities of $\ga 200-400 \, \kms$ (typical of
pulsars), while $3-4 \, \msun$ stars can attain velocities of $\ga
300-400 \, \kms$ (typical of the bound population of halo late
B-type stars). We also found that the ejected stars can occasionally
attain velocities exceeding the Milky Ways's escape velocity.
\end{abstract}


\section{Introduction}
%
The origin of high-velocity runaway stars can be attributed to two
basic processes: (i) disruption of a tight massive binary following
the (asymmetric) supernova explosion of one of the binary components
(Blaauw 1961; Stone 1991; Leonard \& Dewey 1993; Iben \& Tutukov
1996) and (ii) dynamical three- or four-body encounters in dense
stellar systems (Poveda, Ruiz \& Allen 1967; Aarseth 1974; Gies \&
Bolton 1986; Leonard \& Duncan 1990). In the first process, the
maximum velocity attained by runaway stars depends on the magnitude
of the kick imparted to the stellar supernova remnant [either a
neutron star (NS) or a black hole]. For reasonable values of this
magnitude, the runaway velocity does not exceed $\sim 200 \, \kms$
(e.g. Leonard \& Dewey 1993; Portegies Zwart 2000; see also
Gvaramadze 2009). In the second process, the ejection velocity could
be higher. One of the most important and best studied channels for
producing high-velocity stars is through close encounters between
hard (Aarseth \& Hills 1972; Hills 1975; Heggie 1975) binary stars
(Mikkola 1983; Leonard \& Duncan 1990). Numerical simulations by
Leonard (1991) showed that the maximum velocity that a runaway star
(usually the lightest member of the binaries involved in the
interaction) can attain in binary-binary encounters is equal to the
escape velocity from the surface of the most massive star in the
binaries. For binaries containing upper main-sequence stars, the
maximum velocity of runaways could be as large as $\sim 1400 \,
\kms$. The result by Leonard (1991) is often invoked to explain the
high peculiar velocities measured (or inferred) for some runaway
stars (e.g. Heber, Moehler \& Groote 1995; Maitzen et al. 1998;
Tenjes et al. 2001; Ramspeck, Heber, Moehler 2001; Martin 2006;
Gvaramadze 2007; Gvaramadze, Gualandris \& Portegies Zwart 2008,
hereafter Paper\,I; Gvaramadze \& Bomans 2008a; Gvaramadze 2009).

The recent discovery of the so-called hypervelocity stars (HVSs;
Brown et al. 2005; Edelmann et al 2005; Hirsch et al. 2005) -- the
ordinary stars moving with velocities exceeding the Milky Way's
escape velocity -- attracted attention to dynamical processes
involving the supermassive ($\sim 4\times 10^6 \, \msun$) black hole
in the Galactic Centre (Gualandris, Portegies Zwart \& Sipior 2005;
Baumgardt, Gualandris \& Portegies Zwart 2006; Levin 2006; Sesana,
Haardt \& Madau 2006; Ginsburg \& Loeb 2006; Bromley et al. 2006;
Lu, Yu \& Lin 2007; L\"{o}ckmann \& Baumgardt 2008; Perets 2009).
These processes [originally proposed by Hills (1988) and Yu \&
Tremaine (2003)] can result in ejection velocities of several
$1000\kms$. Similar processes but acting in the cores of young
massive star clusters (YMSCs) and involving dynamical encounters
with intermediate-mass ($\sim 100-1000 \, \msun$) black holes
(IMBHs) were considered in Paper\,I (see also Gvaramadze 2006) to
explain the origin of extremely high-velocity ($\ga 1000 \, \kms$)
NSs (e.g. Chatterjee et al. 2005; Hui \& Becker 2006] and HVSs.
Gualandris \& Portegies Zwart (2007) proposed that exchange
encounters between hard binaries and an IMBH formed in the core of a
YMSC in the Large Magellanic Cloud could be responsible for the
origin of the HVS HE\,0437$-$5439 (cf. Edelmann et al 2005;
Przybilla et al. 2008a; Bonanos et al. 2008). A strong support to
the possibility that at least some HVSs originate in star clusters
rather than in the Galactic Centre comes from the proper motion
measurements for the HVS HD\,271791, which constrain the birth place
of this early B-type star to the outer parts of the Galactic disk
(Heber et al. 2008; see also Sect.\,5)\footnote{Note that some
authors (e.g. Brown, Geller \& Kenyon 2009; Tillich et al. 2009)
suggest to use the term ``hypervelocity" to designate the
high-velocity stars ejected solely from the Galactic Centre (i.e.
via the dynamical processes involving the supermassive black hole).
At present, however, proper motion measurements are available for
only one of the known HVSs (see above) so that it is impossible to
unambiguously associate the birthplace of these objects with the
Galactic Centre. In the following we will call ``hypervelocity"
stars all stars with peculiar velocities $\geq 700 \, \kms$.}.
Another example of an extremely high-velocity ($\ga 400 \, \kms$)
B-type star originated in the Galactic disk is the $5 \, \msun$ star
HIP\,60350 (Maitzen et al. 1998), whose birthplace lies at about 7
kpc from the Galactic Centre (Tenjes et al. 2001).

While the existence of a supermassive black hole in the Galactic
Centre is widely accepted, no conclusive evidence has been found for
IMBHs. These objects could be the descendants of very massive ($\ga
1000 \, \msun$) stars (VMSs), formed in the cores of YMSCs through a
runaway sequence of collisions and mergers of ordinary massive stars
(e.g. Portegies Zwart \& McMillan 2002; Portegies Zwart et al. 2004;
see also next Section). Recent numerical studies of the evolution of
VMSs, however, indicate that these stars can lose most of their mass
due to copious stellar winds and leave behind IMBHs with masses of
$\la 70 \, \msun$ (Belkus, Van Bever \& Vanbeveren 2007; Yungelson
et al. 2008), which are not large enough to contribute significantly
to the production of high-velocity runaway stars (see Paper\,I).

In this paper, we explore the hypothesis (Gvaramadze 2007) that some
high-velocity runaway stars could attain their peculiar velocities
in the course of strong dynamical encounters between hard massive
binaries\footnote{We consider a binary as massive if at least one of
its components is more massive than $8 \, \msun$.} and a VMS. In
this process, one of the binary components becomes gravitationally
bound to the VMS, while the second one is ejected, sometimes with a
high velocity. Our goal is to check whether or not this exchange
process can produce early B-type stars (the progenitors of the
majority of NSs) with velocities of $\geq 200-400 \, \kms$ [
measured for some massive halo stars (e.g. Carozzi 1974; Tobin \&
Kaufmann 1984; Keenan et al. 1987; Kilkenny \& Stone 1988; Ramspeck
et al. 2001; Martin 2006) and typical of pulsars (e.g. Hobbs et al.
2005)] and $3-4 \, \msun$ stars with velocities $\geq 300-400 \,
\kms$ (measured for some late B-type halo stars; e.g. Maitzen et al.
1998; Brown et al. 2007). In Section\,2, we discuss the existence of
VMSs in YMSCs and estimate the number of YMSCs formed in the
Galactic disk during a certain interval of time. In Section\,3, we
estimate the typical ejection velocity produced via exchange
encounters between binary stars and a VMS. In Section 4, we compare
this estimate with the results from numerical three-body scattering
experiments. The discussion is given in Section\,5.

\section{Very massive stars}

There is emerging observational evidence that the majority (if not
all) of massive stars form in a cluster environment (e.g.  Zinnecker
\& Yorke 2007 and references therein; cf. de Wit et al.  2005;
Schilbach \& R\"{o}ser 2008; Gvaramadze \& Bomans 2008b) and that
the mass of the most massive star in a cluster is correlated with
the mass of the cluster itself (Elmegreen 2000; Weidner \& Kroupa
2004, 2006). Observations also suggest that the maximum mass of
ordinary stars is saturated at $\sim 150 \, \msun$ for $M_{\rm cl}
\ga 10^5 \, \msun$, where $M_{\rm cl}$ is the mass of the cluster
(Weidner \& Kroupa 2006), a fact which points to the existence of
the upper cut-off of stellar masses (Weidner \& Kroupa 2004; Figer
2005; Oey \& Clarke 2005).

A somewhat higher upper limit on the maximum stellar mass follows
from the recent work by Yungelson et al. (2008), who argue that the
birth masses of some of the observed stars could be as large as
$\simeq 200 \, \msun$ (cf. Oey \& Clarke 2005). It is also possible
that even more massive stars could originate from the coalescence of
binaries whose components have masses close to the upper limit
(Yungelson et al. 2008; cf. Leonard 1995). Moreover, numerical
simulations of the dynamical evolution of YMSCs show that runaway
collisions and mergers of ordinary massive stars can result in the
formation of VMSs with masses as large as $\ga 1000 \, \msun$ (e.g.
Portegies Zwart et al. 1999; Portegies Zwart \& McMillan 2002;
G\"{u}rkan, Freitag \& Rasio 2004; Freitag, G\"{u}rkan \& Rasio
2006; see also Suzuki et al. 2007).

Two necessary conditions should be fulfilled for the runaway growth
of a VMS. First, the parent cluster should be compact (e.g. Gaburov,
Gualandris \& Portegies Zwart 2008; Ardi, Baumgardt \& Mineshige
2008), i.e. dense enough to ensure that stellar collisions are
frequent. Second, the cluster should be massive ($\sim 10^4 -10^5 \,
\msun$), i.e. should contain a large number of OB stars -- the main
building blocks for the VMSs. Let us discuss these conditions in
more detail.

It is conceivable that the first condition is fulfilled for the
majority of YMSCs since their characteristic radius at birth (i.e.
during the embedded phase) is $\la 1$ pc, independently of their
mass (e.g. Kroupa \& Boily 2002 and references therein).  The high
central densities in young clusters could be either primordial (e.g.
Murray \& Lin 1996; Clarke \& Bonnell 2008) or caused by dynamical
mass segregation (e.g.  Portegies Zwart et al. 1999; G\"{u}rkan et
al. 2004). The existing observational data on YMSCs do not allow us
to discriminate between these two possibilities. The observed
top-heavy initial mass functions in the central parts of two of the
most massive and dense YMSCs in our Galaxy, NGC\,3603 and Arches,
could either reflect the initial conditions in the cores of these
clusters (e.g. Harayama, Eisenhauer \& Martins 2008; Dib, Kim \&
Shadmehri 2007; see, however, Ascenso, Alves \& Lago 2008) or have a
dynamical origin (e.g.  Harayama et al.  2008; Portegies Zwart et
al. 2007). It should be noted that the central stellar densities
observed in these two clusters are lower than those required for
runaway collisions to occur. It is possible, however, that either
NGC\,3603 and Arches were collisional at birth and their central
density was already reduced due to dynamical decay or they are
currently on the way to core collapse and runaway collisions. In the
following, we assume that, at a certain stage in the early evolution
of YMSCs, the star density in their cores was high enough for the
runaway collisional process to proceed and for the formation of
VMSs.

Depending on whether the mass segregation in young clusters is
primordial or of dynamical origin, the runaway collisional process
(accompanied by formation of VMSs and ejection of high-velocity
runaway stars) starts respectively at the very beginning of the
cluster evolution or after several Myr (i.e. when the star density in
the cluster's core becomes enhanced via the Spitzer instability). In
the first case, the kinematic age of ejected stars will be close to
the age of the parent cluster, while in the second one, it will be
smaller by several Myr. High-precise proper motion measurements for OB
stars ejected from YMSCs are required to discriminate between these
two possibilities (e.g. Gvaramadze \& Bomans 2008b).

The fact that our Galaxy is populated by YMSCs (the second condition
for the formation of VMSs) was acknowledged only recently (Paper\,I
and references therein; see also Figer 2008). The recent discovery of
several clusters with $M_{\rm cl} \ga 10^4 \, \msun$ (Borissova et
al. 2006; Figer et al. 2006; Davies et al. 2007) and the upward
revision of masses of the already known stellar systems
(Kn\"{o}dlseder 2000; Clark et al. 2005; Homeier \& Alves 2005;
Santos, Bonatto \& Bica 2005; Ascenso et al. 2007a,b; Wolff et
al. 2007; Harayama et al.  2008) indicate that YMSCs are much more
numerous than was hitherto known and that many of them cannot be
detected due to the heavy obscuration caused by the foreground and
circumcluster dusty material.

Below, we estimate the number of star clusters with $M_{\rm cl} \geq
10^4 \, \msun$ formed in the Galactic disk during a certain interval
of time, $\bigtriangleup t$. Since we are interested in the formation
of high-velocity early B-type ($\geq 8 \, \msun$) stars (i.e. the
progenitors of the majority of NSs) and high-velocity late B-type
($\sim 3-4 \, \msun$) stars (representative of the HVSs), we consider
two time intervals, $\bigtriangleup t \sim 30$ Myr (the lifetime of a
single lowest mass star producing a NS) and $\bigtriangleup t \sim
350$ Myr (the lifetime of a $3 \, \msun$ star). Assuming that the star
formation rate of $\sim 7-10\times 10^{-4} \, \msun \, {\rm yr}^{-1}
{\rm kpc}^{-2}$, derived by Lada \& Lada (2003) for embedded star
clusters in the solar neighbourhood, is representative for the
Galactic disk as a whole (see Paper\,I for more details), one has that
$\simeq 6.6-9.4\times 10^6 \, \msun$ and $\simeq 7.7-11.0\times 10^7
\, \msun$ were formed within a circle of 10 kpc during, respectively,
the first and the second time interval. Then taking a power-law
cluster initial mass function with a slope of 2 (e.g. Lada \& Lada
2003) and assuming that the cluster masses range from 50 to $10^6 \,
\msun$, one has that $\sim 70-100$ and $\sim 800-1100$ YMSCs with
$M_{\rm cl} \geq 10^4 \, \msun$ were potentially available to produce
the observed population of high-velocity early and late B-type stars,
respectively. In the following we assume that YMSCs of mass of $\simeq
10^4 \, \msun$ contain ordinary stars as massive as $\simeq 100 \,
\msun$ (Weidner \& Kroupa 2006), while the more massive clusters could
harbour the VMSs formed through the merging of ordinary massive stars.

Similar arguments show that $\sim 10^3$ and $10^4$ star clusters of
mass of $\sim 10^3 \, \msun$ were formed in the Galactic disk
(within a circle of radius of 10 kpc) during the last 30 and 350
Myr, respectively. These clusters produce stars as massive as $50 \,
\msun$ (Elmegreen 2000; Weidner \& Kroupa 2006) and could also
contribute to the production of high-velocity ($\geq 200-400 \,
\kms$) runaway stars (see Section\,3 and Section\,4).

\section{High-velocity stars from exchange encounters between binary stars and a
very massive star}

Close dynamical encounters between (hard) binary stars and a VMS can
result in the ejection of one of the stars with a high velocity. In
this process, one of the binary components is replaced by the VMS,
while the second one is ejected (the so-called exchange
encounter). For equal mass binary components and zero impact
parameter, the typical velocity attained by the escaper is $\sim 1.8
V_{\rm orb}$, where $V_{\rm orb}$ is the orbital velocity of the
ejected star in the original binary (Hills \& Fullerton 1980). The
ejected star gains its kinetic energy at the expense of the increased
binding energy of the post-encounter (newly formed) binary.

The production of high-velocity stars in exchange encounters between
binary stars and a VMS is similar to the process of star ejection in
the course of close dynamical encounters between binaries and a
supermassive black hole (Hills 1988). The main difference is that in
the first process the distance of closest approach of the binary to
the central massive body is limited by the radius of the VMS,
$R_{\rm VMS}$, while in the second one it cannot be smaller than the
tidal radii of the binary components in the gravitational field of
the black hole.  Our goal is to check how this distinction affects
the typical velocity of ejected stars as well as the fraction of
encounters resulting in high-velocity ejections. Below, we use the
results of Hills (1988) to estimate the ejection velocity produced
in exchange encounters and then in Section\,4 we compare this
estimate with results from numerical three-body simulations.

In the process of a close encounter between a binary and a VMS, the
VMS can be treated as a point mass if its radius is smaller than the
binary tidal radius,
\begin{equation}
r_{\rm t} ^{\rm bin} \sim \left({M_{\rm VMS} \over m_1 +m_2 } \right)^{1/3} a \, ,
\label{tid}
\end{equation}
where $M_{\rm VMS}, m_1$ and $m_2$ are, respectively, the masses of
the VMS and the binary components ($m_1 \geq m_2$), and $a$ is the
binary semi-major axis. Evolutionary models of VMSs show that these
stars develop a core-halo configuration (more pronounced at the upper
end of masses), in which most of the mass is concentrated in a dense
and compact core while the rest of the mass is spread in an extended
tenuous halo (Ishii, Ueno, \& Kato 1999; Yungelson et al.
2008). According to Yungelson et al. (2008; also Yungelson, personal
communication), 99 per cent of the mass of 500 and $1000 \, \msun$
VMSs is confined within a sphere of radius $\sim 30$ and $40 \,
\rsun$, respectively. (In Section\,4 we use these figures as input
parameters for our numerical simulations.) For illustrative purposes,
we consider exchange encounters between a VMS and a hard binary
consisting of main-sequence stars with masses $m_1 =40 \, \msun$ and
$m_2 =8 \, \msun$ and semi-major axis $a\simeq 3 r_1 \simeq 30 \,
\rsun$ (typical of tidal binaries; e.g. Lee \& Ostriker 1986), where
\begin{equation}
r_1 =0.8 \left({m_1 \over \msun}\right)^{0.7} R_{\odot} \,
\label{mass-rad}
\end{equation}
is the radius of the primary star (Habets \& Heintze 1981). For these
parameters, one has from equations \,(\ref{tid}) and (\ref{mass-rad})
that $r_{\rm t} ^{\rm bin} \ga 2 R_{\rm VMS}$, so that the VMS can be
treated as a point mass.

The typical ejection velocity at infinity attained by the escapers
in the course of exchange encounters is (Hills 1988)
\begin{eqnarray}
V_{\infty} \simeq 640 \, {\rm km} \, {\rm s}^{-1} \, \alpha
\left({M_{\rm VMS} \over 100 \, \msun} \right)^{1/6}
\left({a \over 30\, R_{\odot}} \right)^{-1/2} \nonumber \\
\times \left[{ m_1 + m_2
\over (40+8)\msun} \right]^{1/3}\, ,
\label{vel}
\end{eqnarray}
where $\alpha$ is a non-monotonic function of the dimensionless
closest approach parameter
\begin{equation}
D_{\rm min} = 79{R_{\rm min} \over r_{\rm t} ^{\rm bin} }
\label{dim}
\end{equation}
and $R_{\rm min}$ is the closest approach distance between the
binary and the VMS; $\alpha \simeq 1$ for $D_{\rm min} \simeq 0$,
then it reaches a maximum of $\simeq 1.2$ for $D_{\rm min} \simeq
30-40$ and then monotonically decreases with increasing $D_{\rm
min}$. For $R_{\rm min} \sim R_{\rm VMS}$ and a VMS of mass $M_{\rm
VMS} = 500$ and $1000 \, \msun$, one has from equation\,(\ref{vel})
and (\ref{dim}) that $D_{\rm min} \simeq 40$ and $V_{\infty}$ is
close to its maximum value of $\simeq 1000$ and $1100 \, \kms$,
respectively. Note that the weak dependence of $V_{\infty}$ on
$M_{\rm VMS}$ implies that exchange encounters with ordinary stars
of mass of $\sim 100 \, \msun$ would in principle be sufficient to
produce HVSs (cf. Section\,4).

\section{Numerical experiments}

In this section, we perform numerical simulations of three-body
encounters in order to obtain the velocity distributions for runaway
stars produced in the course of exchange interactions between hard
binary stars and a massive compact body, either a VMS or an ordinary
star of mass of $50-100 \, \msun$. The simulations are carried out
using the {\tt sigma3} package, which is part of the
STARLAB\footnote{\tt http://www.manybody.org/manybody/starlab.html}
software environment (McMillan \& Hut 1996; Portegies Zwart et al.
2001). For a detailed description of the setup of the scattering
experiments see Gualandris et al. (2005). During the simulations we
allow for physical collisions when the distance between any two
stars is smaller than the sum of their radii. For ordinary stars we
use the mass-radius relationship given by
equation\,(\ref{mass-rad}), while for the VMSs of mass of $500 \,
\msun$ and $1000 \, \msun$ we adopt the radii $R_{\rm VMS} = 30 \,
\rsun$ and $40 \, \rsun$, respectively (see Section\,3). The initial
eccentricity of the stellar binary is randomly drawn from a thermal
distribution in the allowed range $(0-e_{\rm max})$, where $e_{\rm
max} = 1 - 2 (r_1+r_2)/a$ is chosen as to avoid a collision at the
first pericentre passage ($r_2$ is the radius of the secondary
star). The relative velocity at infinity between the centre of mass
of the binary and the central massive body is set to $5\, \kms$, in
accordance with typical dispersion velocities in YMSCs.

\subsection{High-velocity early B-type stars}

First, we focus on exchange encounters producing high-velocity early
B-type stars. From numerical three-body scattering experiments, it
is known that the least massive binary component is more likely to
be ejected with high velocity if its companion star is much more
massive and/or the binary semi-major axis is small (e.g. Gualandris
\& Portegies Zwart 2007). For illustrative purposes we consider
encounters between a binary consisting of two main-sequence stars
with masses $m_1 = 40 \, \msun$ and $m_2 =8\, \msun$  and a (very)
massive star with $M_{\rm VMS}$ ranging from 50 to $1000 \, \msun$.
In order to maximize the ejection speed, we assume that the binary
system is very tight, e.g. formed by tidal capture. In this case,
$a\simeq 3 \, r_1 \simeq 0.15$ AU (see Section\,3 and cf. Paper\,I).

\begin{figure}
\begin{center}
\includegraphics[width=8cm]{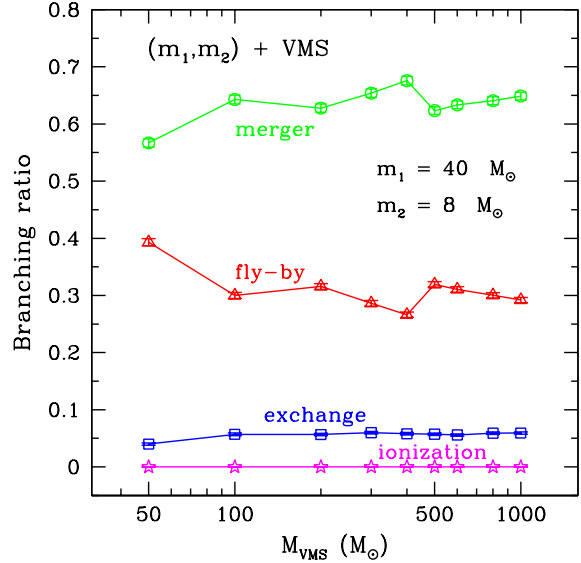}
\end{center}
\caption{Branching ratio for the outcome of encounters between a
  $(40,8)\, \msun$ binary and a single VMS as a function of the VMS mass.
  The different outcomes are: merger (circles), fly-by (triangles),
  ionization (stars) and exchange (squares). The error bars represent
  the formal ($1\sigma$) Poissonian uncertainty of the measurement.}
\label{fig:branch1}
\end{figure}

In Fig.\,\ref{fig:branch1} we present the probability of different
outcomes (branching ratios) as a function of $M_{\rm VMS}$. For each
value of $M_{\rm VMS}$ we perform a total of 10000 scattering
experiments, which result either in a fly-by, an exchange or a
merger. Ionizations never take place as the binary is too hard to be
dissociated by the VMS. Mergers occur in a large fraction ($\sim$65
per cent) of encounters due to the small semi-major axis of the
binary and the finite size of the stars. Exchange interactions occur
in about 6 per cent of encounters. During these interactions, one of
the binary components is captured by the VMS while the other is
ejected, sometimes with a high velocity. These encounters are the
relevant ones for the production of high-velocity runaway stars.

\begin{figure}
\begin{center}
\includegraphics[width=8cm]{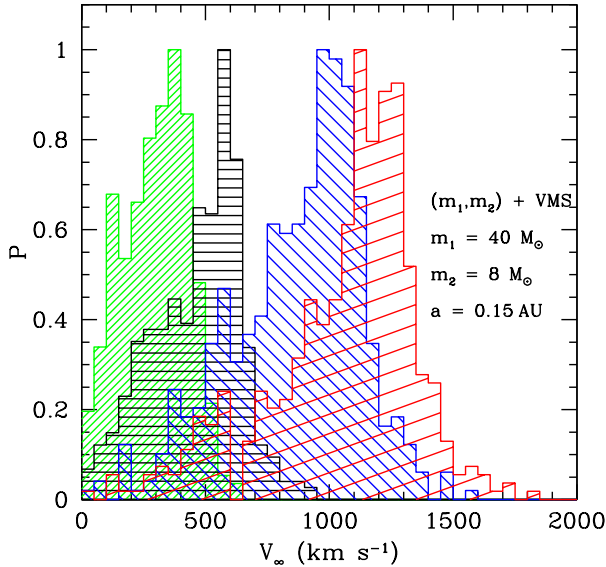}
\end{center}
\caption{Velocity distributions at infinity for escaping ($8 \, \msun$) stars in encounters
  between a binary consisting of a primary star with mass $m_1 =
  40\, \msun$ and a secondary star with mass $m_2 = 8\, \msun$, and
  a single VMS of mass $M_{\rm VMS}= 50\, \msun , 100\, \msun , 500\, \msun$
  and $1000\, \msun$ (from left to right). The binary semi-major axis is $a$ =
  0.15 \,AU.}
\label{fig:vel1}
\end{figure}

Fig.\,\ref{fig:vel1} shows the velocity distribution for $8 \,
\msun$ escapers. As expected, the more massive VMSs are more likely
to eject stars with high velocities. Fig.\,\ref{fig:vel1} also shows
that the typical ejection velocities attained by the escapers are
consistent with the predictions derived in Section\,3.
\begin{figure}
\begin{center}
\includegraphics[width=8cm]{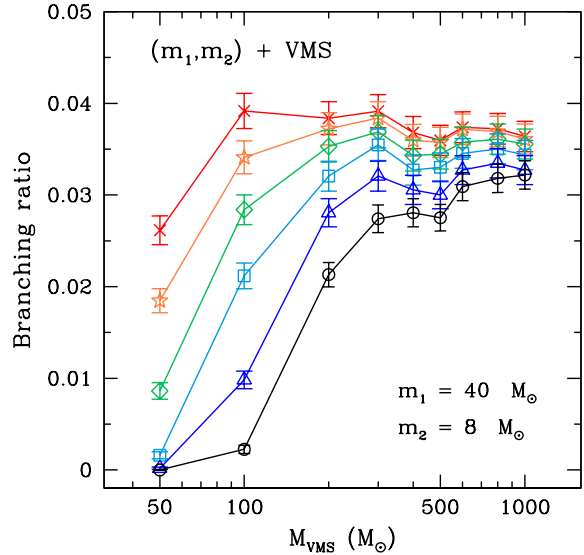}
\end{center}
\caption{The probability of exchange encounters between a $(40,8)\,
\msun$ binary (with $a=0.15$ AU) and a single VMS resulting in ejection
of the $8 \, \msun$ star with a velocity from 200 to $700 \, \kms$ (
top to bottom).}
\label{fig:branchm1}
\end{figure}
In Fig.\,\ref{fig:branchm1} we show the probability of exchange
encounters resulting in ejection of the $8 \, \msun$ binary component
with a velocity from 200 to $700 \, \kms$ (top to bottom).  For
$M_{\rm VMS} \ga 100 \, \msun$ about $3$ per cent of all encounters
produce runaways with peculiar velocities $\geq 400 \, \kms$. It can
be seen (see also Fig.\,\ref{fig:vel1}) that even an ordinary star of
mass of $50 \, \msun$ can occasionally (in $\sim 1$ per cent of
encounters) produce an escape velocity $\geq 400 \, \kms$. In order to
produce escapers with velocities typical of HVSs ($V_{\infty} \geq 700
\, \kms$), a VMS of several hundred solar masses is required. In the
case of a $200-300 \, \msun$ VMS, $\ga 2$ per cent of all encounters
result in an escape velocity of $\geq 700 \, \kms$. This fraction
gradually increases to $\ga 3$ per cent for the more massive VMSs.

\begin{figure}
\begin{center}
\includegraphics[width=8cm]{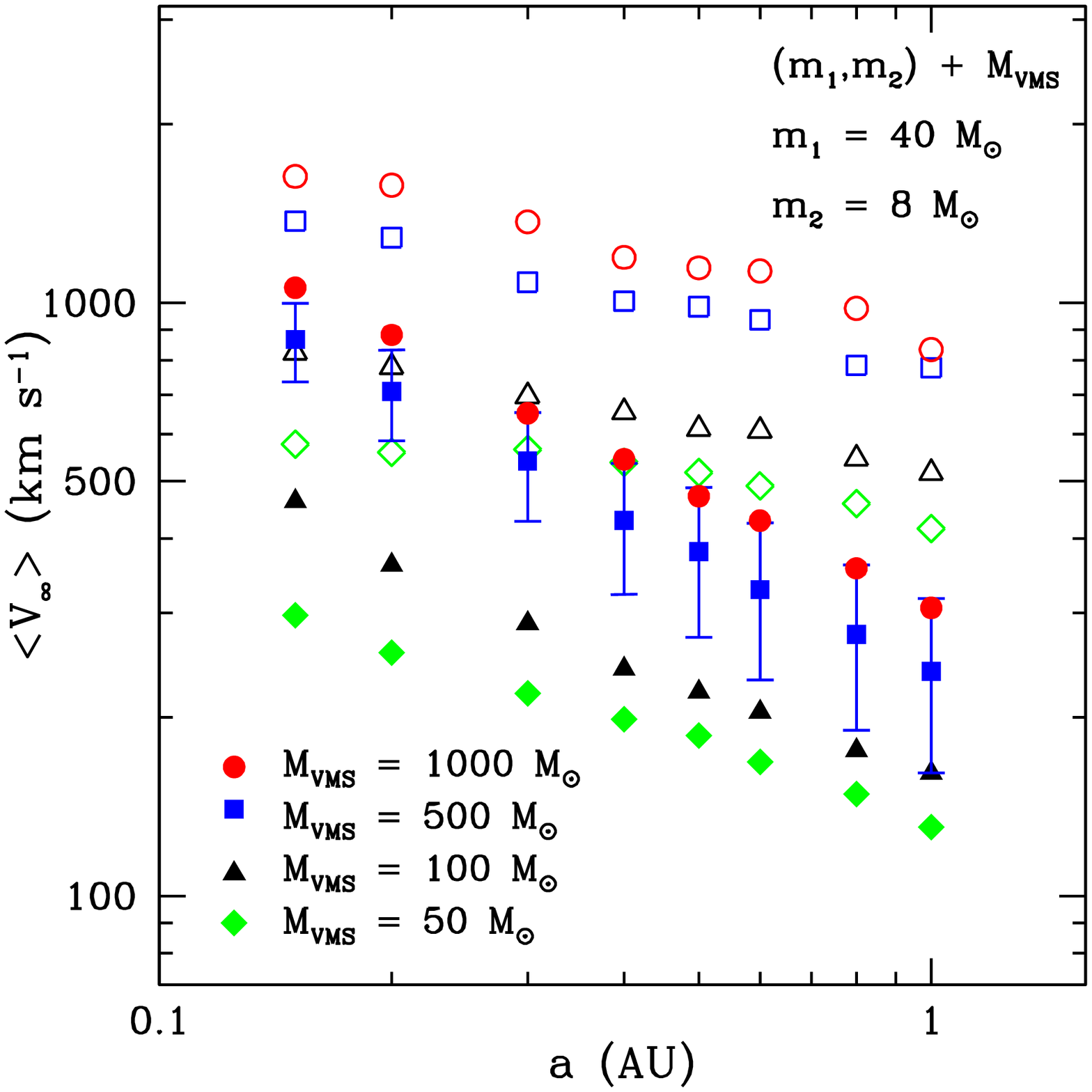}
\end{center}
\caption{Average recoil velocity of escapers as a function of the
  initial binary semi-major axis in the interaction of a $(40,8)\,
  \msun$ binary star with VMSs of different mass: $M_{\rm VMS} =50\,
  \msun$ (diamonds), $M_{\rm VMS} =100\, \msun$ (triangles), $M_{\rm
    VMS} =500\, \msun$ (squares), $M_{\rm VMS} =1000\, \msun$
  (circles). Solid symbols represent the average velocity obtained
  from a set of 10000 scattering experiments while the empty symbols
  indicate the velocity $V_{\rm max}$ for which 1 per cent of the
  escapers have $V_{\infty} > V_{\rm max}$. The error bars indicate
  the 1$\sigma$ deviation from the mean. For clarity, we only show
  them for one data set.}
\label{fig:velo1}
\end{figure}
Fig.\,\ref{fig:velo1} shows the average recoil velocity, $\langle
V_{\infty} \rangle$, of escapers ($8 \, \msun$ stars; solid symbols)
as a function of the binary semi-major axis for four different
values of the VMS mass $M_{\rm VMS} =50\, \msun , 100\, \msun ,
500\, \msun$ and $1000\, \msun$. The empty symbols indicate the
velocity $V_{\rm max}$ for which 1 per cent of the escapers have
$V_{\infty} > V_{\rm max}$. The average and the maximum velocities
increase with the mass of the VMS, as expected from energetic
arguments. On the other hand, the fraction of high-velocity escapers
decreases rapidly with increasing $a$. For $M_{\rm VMS} =50-100 \,
\msun$, escapers are ejected with velocities $\geq 200-400 \, \kms$
if the binary semi-major axis is in the range 0.15 AU $<a<0.6$ AU.
The same velocities could be achieved with wider binaries (with $a$
up to 1 AU) if $M_{\rm VMS} \ga 300 \, \msun$. To produce
hypervelocity escapers $M_{\rm VMS}$ should be $\ga 500 \, \msun$
and the smallest possible semi-major axes are needed ($a\simeq
0.15-0.2$ AU). Fig.\,\ref{fig:velo1} also shows that even a $100 \,
\msun$ star can occasionally produce hypervelocity escapers, but the
fraction of these events is very small (see also
Fig.\,\ref{fig:branchm1}).

Note that the average recoil velocity of escapers produced in
exchange encounters with a VMS is somewhat larger than that produced
in encounters with an IMBH of the same mass (cf.
Fig.\,\ref{fig:velo1} with Fig.\,4 in Paper\,I). This seemingly
``incorrect" result can be understood if one takes into account that
the ejection velocity is a non-monotonic function of the closest
approach of the binary to the central massive body (either a VMS or
an IMBH) and that for encounters with the VMS $R_{\rm min}$ cannot
be less than $R_{\rm VMS}$, while for encounters with the IMBH
$R_{\rm min}$ is limited by the tidal radius of the ejected star,
$r_{\rm t} ^{\star} \sim (M_{\rm IMBH} /m_{\star} )^{1/3}
r_{\star}$, where $M_{\rm IMBH}$ is the mass of the IMBH and
$m_{\star}$ and $r_{\star}$ are the mass and the radius of the star.
For the parameters adopted in our simulations, one has from
equations\,(\ref{tid}) and (\ref{dim}) that $D_{\rm min} \simeq
40-50$ and $\simeq 15$ for encounters with the VMSs and the IMBHs,
respectively. Thus, in the first process the smaller $R_{\rm min}$
the larger $V_{\infty}$, while in the second one the closest
possible approach to the black hole does not produce the maximum
ejection velocity (see Section\,3; see also Fig.\,4 in Gualandris \&
Portegies Zwart 2007)), which in turn results in the smaller
$\langle V_{\infty} \rangle$.

\subsection{High-velocity late B-type stars}

\begin{figure}
\begin{center}
\includegraphics[width=8cm]{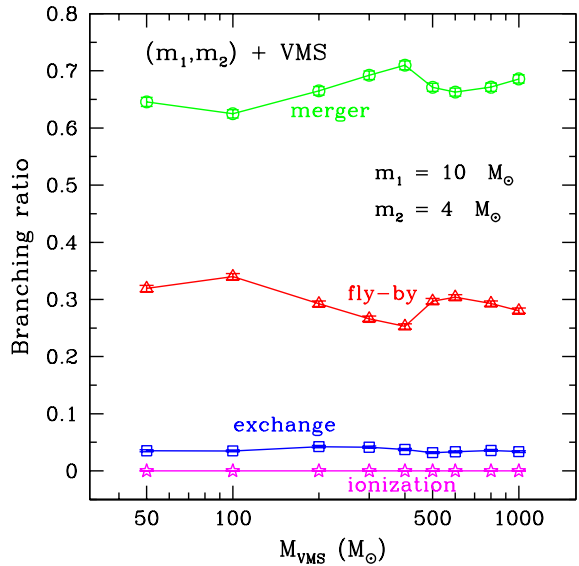}
\end{center}
\caption{Branching ratio for the outcome of encounters between a
  $(10,4)\, \msun$ binary and a single VMS as a function of the VMS mass.
  The different outcomes are: merger (circles), fly-by (triangles),
  ionization (stars) and exchange (squares). The error bars represent
  the formal ($1\sigma$) Poissonian uncertainty of the measurement.}
\label{fig:branch2}
\end{figure}

Now we simulate exchange encounters producing high-velocity late
B-type stars. In order to derive the probability of obtaining the
largest possible ejection velocities, we consider encounters between a
very tight binary ($m_1 =10 \, \msun$, $m_2 =4 \, \msun$ and $a\simeq
3 r_1 \simeq 0.06$ AU) and a (very) massive star of mass of $50-1000
\, \msun$. The branching ratio for these encounters
(Fig.\,\ref{fig:branch2}) is almost identical to that given in
Fig.\,\ref{fig:branch1}. The only difference is the somewhat smaller
percentage of exchange encounters due to the smaller semi-major axis
of the binary.

\begin{figure}
\begin{center}
\includegraphics[width=8cm]{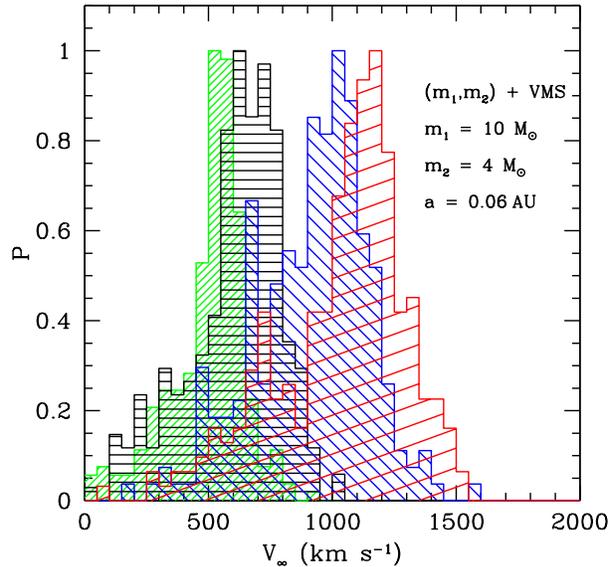}
\end{center}
\caption{Velocity distributions at infinity for escaping stars in encounters
  between a binary consisting of a primary star with mass $m_1 =
  10\, \msun$ and a secondary star with mass $m_2 = 4\, \msun$, and
  a single VMS of mass $M_{\rm VMS}=50\, \msun$, $M_{\rm VMS}=100\, \msun$,
  $500\, \msun$ and $1000\, \msun$ (from left to right). The binary semi-major
  axis is $a$ = 0.06 \,AU.}
\label{fig:velhisto2}
\end{figure}

\begin{figure}
\begin{center}
\includegraphics[width=8cm]{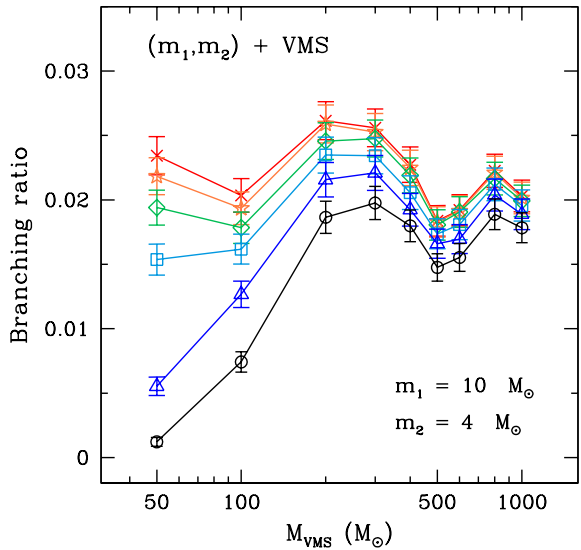}
\end{center}
\caption{The probability of exchange encounters between a $(10,4)\,
  \msun$ binary (with $a=0.06$ AU) and a single VMS resulting in ejection
  of the $4 \, \msun$ star with a velocity from 200 to $700 \, \kms$
  (top to bottom).}
\label{fig:branchm2}
\end{figure}

The velocity distributions for $4 \, \msun$ escapers are shown in
Fig.\,\ref{fig:velhisto2} for four different values of the VMS mass:
$M_{\rm VMS}= 50, 100, 500$ and $1000 \, \msun$. The typical
ejection velocity obtained for each set of parameters is consistent
with the estimates derived from equation\,(\ref{vel}).
Fig.\,\ref{fig:branchm2} shows that for all values of $M_{\rm VMS}$
about $2$ per cent of encounters results in peculiar velocities
$\geq 300-400 \, \kms$ and that about the same percentage of ejected
stars attains velocity $\geq 700 \, \kms$ if $M_{\rm VMS} \ga 200 \,
\msun$.

\begin{figure}
\begin{center}
\includegraphics[width=8cm]{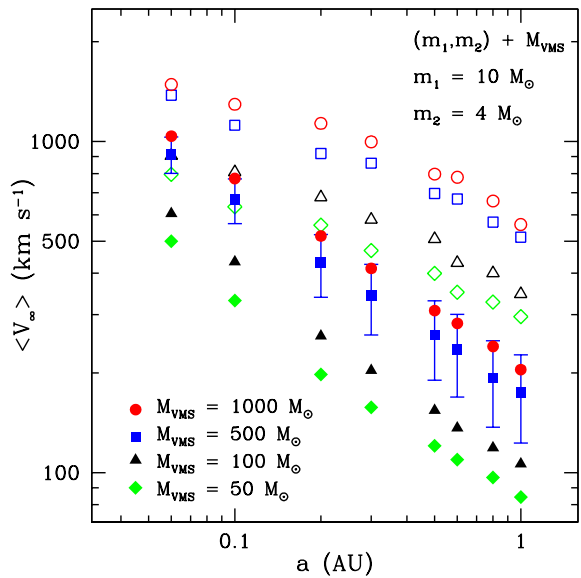}
\end{center}
\caption{Average recoil velocity of escapers as a function of the
  initial binary semi-major axis in the interaction of a $(10,4)\,
  M_{\odot}$ binary star with VMSs of different mass: $M_{\rm VMS}
  =50\, \msun$ (diamonds), $M_{\rm VMS} =100\, \msun$ (triangles),
  $M_{\rm VMS} =500\, \msun$ (squares), $M_{\rm VMS} =1000\, \msun$
  (circles). Solid symbols represent the average velocity obtained
  from a set of 10000 scattering experiments while the empty symbols
  indicate the velocity $V_{\rm max}$ for which 1 per cent of the
  encounters have $V_{\infty} > V_{\rm max}$. The error bars indicate
  the 1$\sigma$ deviation from the mean. For clarity, we only show
  them for one data set.}
\label{fig:velo2}
\end{figure}

\begin{figure}
\begin{center}
\includegraphics[width=8cm]{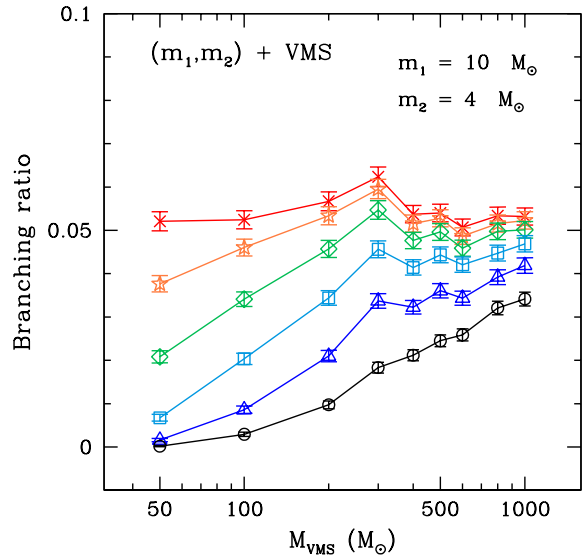}
\end{center}
\caption{The same as Fig.\,\ref{fig:branchm2} but for binaries with
 semi-major axis $a=0.1$ AU.}
\label{fig:branchm2m}
\end{figure}

The average recoil velocity of escapers as a function of the binary
semi-major axis is shown in Fig.\,\ref{fig:velo2}. As in the case of
encounters between $(40+8)\msun$ binaries and a VMS, the average and
the maximum velocities increase with the mass of the VMS. Both
velocities, however, can reach somewhat higher values due to the
smaller possible semi-major axis of the binary (cf. with
Fig.\,\ref{fig:velo1}).
Note also that although $\langle V_{\infty} \rangle$ decreases with
increase of $a$, the fraction of escapers with a given velocity ($<
\langle V_{\infty} \rangle$) could be larger for wider binaries
(this is because the percentage of encounters resulting in exchanges
increases with increase of $a$; see e.g. Fig.\,1 in Gualandris \&
Portegies Zwart 2007). This effect is illustrated in
Fig.\,\ref{fig:branchm2m}, which shows that for binaries with
$a=0.1$ AU the percentage of escapers with $V_{\infty} \geq 400 \,
\kms$ is about 2.5 times larger as compared to the case of the more
tight binaries (cf. Fig.\,\ref{fig:branchm2}).

\section{Discussion}

We performed numerical simulations of dynamical encounters between
hard massive binaries and a VMS, in order to explore the hypothesis
that this dynamical process could be responsible for the origin of
high-velocity ($\ga 200-400 \, {\rm km} \, {\rm s}^{-1}$) runaway
stars. In our study we proceeded from the similarity between
encounters involving a VMS and those involving an IMBH (the latter
process is already known to be able to produce high-velocity
runaways; Paper\,I; Gualandris \& Portegies Zwart 2007) and the fact
that the radii of VMSs are smaller than the tidal radii of the
intruders (the massive binaries), so that the tidal breakup and
ejection can occur before the binary components merge with the VMS
(Section\,3; see also Gvaramadze 2007). Our study was motivated by
the recent evolutionary models of VMSs (Belkus et al. 2007;
Yungelson et al. 2008), which suggest that VMSs can lose most of
their mass via copious winds and leave behind IMBHs with masses of
$\la 70 \, \msun$, which are not large enough to contribute
significantly to the production of high-velocity runaway stars (see
Paper\,I). We therefore explored the possibility that a VMS could
produce high-velocity escapers (either early or late B-type stars)
before it finished its life in a supernova and formed a black hole.
We estimated the typical velocities produced in encounters between
very tight massive binaries and VMSs (with $M_{\rm VMS} \ga 200 \,
\msun$) and found that about $3-4$ per cent of all encounters
produce velocities of $\geq 400 \, \kms$, while in about 2 per cent
of encounters the escapers attain velocities comparable to those
measured for HVSs (i.e. $\geq 700 \, \kms$). We therefore argue that
the origin of high-velocity ($\geq 200-400 \, \kms$) runaway stars
and at least some HVSs could be associated with dynamical encounters
between the tightest massive binaries and VMSs formed in the cores
of YMSCs. In this connection, it is worthy to note that the
theoretical velocity distribution for HVSs produced in the Galactic
Centre (i.e. through the dynamical processes involving the
supermassive black hole) predicts the existence of a tail of
velocities of up to several thousand $\kms$, while the peculiar
velocities observed for all known ($\sim 20$) HVSs do not exceed
$\sim 800 \, \kms$ (e.g. Sesana, Haardt \& Madau 2007; L\"{o}ckmann
\& Baumgardt 2008). Interestingly, the latter figure better agrees
with the maximum ejection velocity ($\simeq 1000-1500 \, \kms$)
produced via dynamical three- and four-body encounters in dense star
clusters (see Section\,3 and Paper\,I). The future proper motion
measurements for HVSs with GAIA will reveal what fraction of these
extremely high-velocity stars originated in the Galactic disk.

We also simulated dynamical encounters between tight massive
binaries and single $50-100\, \msun$ stars -- the most massive
ordinary stars formed in clusters with $M_{\rm cl} \simeq 10^3 -10^4
\, \msun$ (Weidner \& Kroupa 2006). We found that from 1 to $\simeq
4$ per cent of these encounters can produce runaway stars with
velocities of $\geq 300-400 \, \kms$ (typical of the bound
population of high-velocity halo B-type stars) and occasionally (in
less than 1 per cent of encounters) produce hypervelocity ($\geq 700
\, \kms$) late B-type escapers. Note that the smaller production
rate of high-velocity escapers in this case could be compensated by
an order of magnitude larger population of clusters containing $50
\, \msun$ stars (see Section\,2).

Our explanation for the origin of high-velocity runaway stars
requires a very dense stellar environment of the order of $10^6
-10^7$ stars ${\rm pc}^{-3}$ (see Paper\,I). The role of this
environment is three-fold. First, it makes possible the runaway
merging process, resulting in the formation of VMSs. Second, it
provides suitable conditions for production of tight binaries via
the tidal capture process and hardens the existing binaries, thereby
increasing the probability of energetic three-body encounters.
Third, it ensures that the three-body dynamical encounters are
frequent, i.e. the production rate of high-velocity escapers could
be high. We caution, however, that whether or not such high
densities exist in the cores of star clusters remains unclear to
date (see Section\,2 and cf. Paper\,I). This and numerous
uncertainties about the initial conditions and early evolution of
young star clusters precludes us from making any estimates of the
production rate of high-velocity escapers (cf. Paper\,1).

In Paper\,1, we suggested that some extremely high-velocity NSs
could be the remnants of hypervelocity massive stars ejected via
strong dynamical three- or four-body interactions in the cores of
YMSCs (i.e. the origin of these NSs should not necessarily be
connected with asymmetric supernova explosions). Our suggestion was
based on the fact that one of the HVSs known at that time,
HE\,0437$-$5439, is massive enough ($\simeq 9 \, M_{\odot}$;
Przybilla et al. 2008a) to explode as a type\,II supernova and to
leave behind a high-velocity NS. It is important to note that the
time of flight of this star from the Galactic Centre exceeds the
lifetime of the star and therefore the extremely high peculiar
velocity of HE\,0437$-$5439 cannot be attributed to the dynamical
processes involving the supermassive black hole in the Galactic
Centre. The most likely birth place of this HVS is one of the YMSCs
in the Large Magellanic Cloud (Gualandris \& Portegies Zwart 2007;
cf. Przybilla et al. 2008a; Bonanos et al. 2008; see however Perets
2008). Recently, an even more massive ($\simeq 11\pm 1 \, \msun$)
HVS, HD\,271791, was discovered (Heber et al. 2008; see also Carozzi
1974; Kilkenny \& Stone 1988) whose birth place cannot be associated
with the Galactic Centre. HD\,271791 is the only HVS with measured
proper motion and all measurements indicate that this star
originated in the periphery of the Galactic disk (Heber et al.
2008). Thus, the high peculiar velocity of HD\,271791 cannot be
attributed to the ejection mechanism involving the supermassive
black hole in the Galactic Centre. Przybilla et al. (2008b) proposed
that HD\,271791 was a member of a massive close binary system
disrupted in an asymmetric supernova explosion and that the
secondary star (HD\,271791) was released at its orbital velocity
($\sim 400 \, \kms$) in the direction of Galactic rotation (which
allowed them to explain the Galactic rest-frame velocity of
HD\,271791, provided that it is on the low end of the observed range
$530-920 \, \kms$).  One can show, however, that to explain the high
space velocity of HD\,271791 within the framework of the
binary-supernova scenario, the stellar remnant of the supernova
explosion [a $\la 10 \, \msun$ black hole, according to Przybilla et
al. (2008b)] should receive at birth an unrealistically large kick
velocity of $750-1200 \, \kms$ (Gvaramadze 2009). We therefore
believe that the more likely origin of the peculiar velocity of
HD\,271791 (and other halo early B-type stars; e.g. Ramspeck et al.
2001; Martin 2006) is through the dynamical processes discussed in
the present paper and in Paper\,I.

The same dynamical processes could also be responsible for the origin
of early B-type stars observed in the halos of nearby galaxies
(e.g. Comer\'{o}n, G\'{o}mez \& Torra 2003). We speculate that these
stars can meet and ionize the cloudlets of cold gas on their way
through the halo and suggest that the so-called extraplanar HII
regions (e.g. T\"{u}llmann et al. 2003) could be the Str\"{o}mgren
zones produced by high-velocity runaway OB stars (cf.  Gvaramadze \&
Bomans 2008a). Our suggestion could be supported by the fact that the
${\rm H}_\alpha$-luminosities of two extraplanar HII regions in the
galaxy NGC\,55 (located at $\sim 0.8$ and 1.5 kpc from the galactic
plane) are consistent with the possibility that the ionizing sources
of these objects are single B0 or O9.5 stars (T\"{u}llmann et
al. 2003).

High-velocity early B-type stars ejected at large angles to the
Galactic plane end their lives in the halo and thereby contribute to
the population of halo NSs. A possible example of a NS formed in the
halo is the high Galactic latitude compact X-ray source 1RXS
J141256.0+792204. This object has a high X-ray to optical flux ratio
($> 8700$), typical of isolated NSs (Rutledge, Fox \& Shevchuk
2008). If 1RXS J141256.0+792204 is indeed an isolated NS, then its
distance from the Sun is $\simeq 8.4$ kpc, that corresponds to the
distance $z\simeq 5.1$ kpc from the Galactic plane (Rutledge et al.
2008). The surface temperature of 1RXS J141256.0+792204 (inferred
from modelling its spectrum as thermal blackbody) suggests that this
NS should be $\la 10^6$ yr old, provided that its cooling follows
the standard cooling curves. If one assumes that the NS was born
near the Galactic plane, then its peculiar velocity should be $>
5000 \, {\rm km} \, {\rm s}^{-1}$, which is too high to be
realistic.  To avoid this problem, Rutledge et al. (2008) suggested
that either 1RXS J141256.0+792204 is not an isolated NS (i.e. the
object is much closer to the Sun) or its cooling is non-standard
(i.e. the NS is much older). Instead, we suggest that 1RXS
J141256.0+792204 could be the remnant of a supernova explosion of an
early B-type star ejected from a YMSC in the Galactic disk with a
velocity of $z/t_{\ast} \ga 170\, \kms$, where $t_{\ast} \la 30$ Myr
is the lifetime of the star. It is obvious that even in the case of
a symmetric supernova explosion the peculiar velocity of the stellar
remnant (the NS) will be as large as that of its progenitor star
(Paper\,I). We argue therefore that the proper motion of 1RXS
J141256.0+792204 could be as small as $\sim 4 \, {\rm mas} \, {\rm
yr}^{-1}$ [i.e. $\sim 20$ times less than that suggested by Rutledge
et al. (2008)] or even smaller if the NS is near the apex of its
trajectory. In the latter case, the direction of the proper motion
of the NS could be arbitrary.

\section{Acknowledgements}
We are grateful to L.R.Yungelson for useful discussions. VVG
acknowledges the Deutsche Forschungsgemeinschaft and the Deutscher
Akademischer Austausch Dienst for partial financial support. AG is
supported by grant NNX07AH15G from NASA. SPZ acknowledges support
from the Netherlands Organization for Scientific Research (NWO under
grant No. 643.200.503) and the Netherlands Research School for
Astronomy (NOVA).


\begin{thebibliography}{}
%
\bibitem{} Aarseth S.J., 1974, A\&A, 35, 237
\bibitem{} Aarseth S.J., Hills J.G., 1972, A\&A, 21, 255
\bibitem{} Ardi E., Baumgardt H., Mineshige S., 2008, ApJ, 682, 1195
\bibitem{} Ascenso J., Alves J., Lago M.T.V.T., 2008, preprint(astro-ph/0811.3213)
\bibitem{} Ascenso J., Alves J., Beletsky Y., Lago M.T.V.T., 2007a, A\&A, 466, 137
\bibitem{} Ascenso J., Alves J., Vicente S., Lago M.T.V.T., 2007b, A\&A, 476, 199
\bibitem{} Baumgardt H., Gualandris A., Portegies Zwart S., 2006, MNRAS, 372, 174
\bibitem{} Belkus H., Van Bever J., Vanbeveren D., 2007, ApJ, 659, 1576
\bibitem{} Blaauw A., 1961, Bull. Astron. Inst. Netherlands, 15, 265
\bibitem{} Bonanos A.Z., L\'{o}pez-Morales M., Hunter I., Ryans R.S.I., 2008, ApJ, 675, L77
\bibitem{} Borissova J., Ivanov V.D., Minniti D., Ceisler D., 2006, A\&A, 455, 923
\bibitem{} Bromley B.C. et al., 2006, ApJ, 653, 1194
\bibitem{} Brown W.R., Geller M.J., Kenyon S.J., Kurtz M.J., 2005, ApJ, 622, L33
\bibitem{} Brown W.R., Geller M.J., Kenyon S.J., Kurtz M.J, Bromley B.C., 2007, ApJ, 660, 311
\bibitem{} Brown W.R., Geller M.J., Kenyon S.J., 2009, ApJ, 690, 1639
\bibitem{} Carozzi N., 1974, A\&AS, 16, 277
\bibitem{} Chatterjee S. et al., 2005, ApJ, 630, L61
\bibitem{} Clark J.S., Negueruela I., Crowther P.A., Goodwin S.P., 2005, A\&A, 434, 949
\bibitem{} Clarke C.J, Bonnell I.A., 2008, MNRAS, 388, 1171
\bibitem{} Comer\'{o}n F., G\'{o}mez A.E., Torra J., 2003, A\&A, 400, 137
\bibitem{} Davies B., Figer D.F., Kudritzki R.-P., MacKenty J., Najarro F., Herrero A., 2007, ApJ, 671, 781
\bibitem{} de Wit W.J., Testi L., Palla F., Zinnecker H., 2005, A\&A, 437, 247
\bibitem{} Dib S., Kim J., Shadmehri M., 2007, MNRAS, 381, L40
\bibitem{} Edelmann H., Napiwotzki R., Heber U., Christlieb N., Reimers D., 2005, ApJ, 634, L181
\bibitem{} Elmegreen B.G., 2000, ApJ, 539, 342
\bibitem{} Figer D.F., 2005, Nat, 434, 192
\bibitem{} Figer D.F., 2008, in ``Massive Stars as Cosmic Engines", IAU Symp. 250, ed. F. Bresolin, P. A. Crowther, \& J. Puls
(Cambridge Univ. Press), p. 247
\bibitem{} Figer D.F., MacKenty J., Robberto M., Smith K., Najarro F., Kudritzki R.P., Herrero A., 2006, ApJ, 643, 1166
\bibitem{} Freitag M., G\"{u}rkan M.A., Rasio F.A., 2006, MNRAS, 368, 141
\bibitem{} Gaburov E., Gualandris A., Portegies Zwart S., 2008, MNRAS, 384, 376
\bibitem{} Gies D.R., Bolton C.T., 1986, ApJS, 61, 419
\bibitem{} Ginsburg I., Loeb A., 2006, MNRAS, 368, 221
\bibitem{} Gualandris A., Portegies Zwart S., 2007, MNRAS, 376, L29
\bibitem{} Gualandris A., Portegies Zwart S., Sipior M.S., 2005, MNRAS, 363, 223
\bibitem{} G\"urkan M.A., Freitag M., Rasio F.A., 2004, ApJ, 604, 632
\bibitem{} Gvaramadze V.V., 2006, in On the Present and Future of Pulsar Astronomy, 26th meeting
of the IAU, Joint Discussion 2, 16-17 August, 2006, Prague, Czech
Republic, JD02, \# 25
\bibitem{} Gvaramadze V.V., 2007, A\&A, 470, L9
\bibitem{} Gvaramadze V.V., 2009, MNRAS, in press (astro-ph/0902.4815)
\bibitem{} Gvaramadze V.V., Bomans D.J., 2008a, A\&A, 485, L29
\bibitem{} Gvaramadze V.V., Bomans D.J., 2008b, A\&A, 490, 1071
\bibitem{} Gvaramadze V.V., Gualandris A., Portegies Zwart S., 2008, MNRAS, 385,
929 (Paper\,I)
\bibitem{} Habets G.M.H.J., Heintze J.R.W., 1981, A\&AS, 46, 193
\bibitem{} Harayama Y., Eisenhauer F., Martins F., 2008, ApJ, 675, 1319
\bibitem{} Heber U., Moehler S., Groote D., 1995, A\&A, 303, L33
\bibitem{} Heber U., Edelmann H., Napiwotzki R., Altmann M., Scholz R.-D., 2008, A\&A, 483, L21
\bibitem{} Heggie D.C., 1975, MNRAS, 173, 729
\bibitem{} Hills J.G., 1975, AJ, 80, 809
\bibitem{} Hills J.G., 1988, Nat, 331, 687
\bibitem{} Hills J.G., Fullerton, L.W., 1980, AJ, 85, 1281
\bibitem{} Hirsch H.A., Heber U., O'Toole S. J., Bresolin F., 2005, A\&A, 444, L61
\bibitem{} Hobbs G., Lorimer D.R., Lyne A.G., Kramer M., 2005, MNRAS, 360, 974
\bibitem{} Homeier N., Alves J., 2005, A\&A, 430, 481
\bibitem{} Hui C.Y, Becker W., 2006, A\&A, 457, L33
\bibitem{} Iben I., Jr., Tutukov A.V., 1996, ApJ, 456, 738
\bibitem{} Ishii M., Ueno M., Kato M., 1999, PASJ, 51, 417
\bibitem{} Keenan F.P., Brown P.J.F., Conlon E.S., Dufton P.L., Lennon
D.J., 1987, A\&A, 178, 194
\bibitem{} Kilkenny D., Stone, L.E., 1988, MNRAS, 234, 1011
\bibitem{} Kn\"{o}dlseder J., 2000, A\&A, 360, 539
\bibitem{} Kroupa P., Boily C.M., 2002, MNRAS, 336, 1188
\bibitem{} Lada C.J., Lada E.A., 2003, ARA\&A, 41, 57
\bibitem{} Lee H.M., Ostriker J.P., 1986, ApJ, 310, 176
\bibitem{} Leonard P.J.T., 1991, AJ, 101, 562
\bibitem{} Leonard P.J.T., 1995, MNRAS, 277, 1080
\bibitem{} Leonard P.J.T., Dewey R.J., 1993, in Luminous High-Latitude Stars, ed. D.D. Sasselov (San Francisco: ASP), 239
\bibitem{} Leonard P.J.T., Duncan M.J., 1990, AJ, 99, 608
\bibitem{} Levin Y., 2006, ApJ, 653, 1203
\bibitem{} L\"{o}ckmann U., Baumgardt H., 2008, MNRAS, 384, 323
\bibitem{} Lu Y., Yu Q., Lin D.N.C., 2007, ApJ, 666, L89
\bibitem{} Maitzen H.M., Paunzen E., Pressberger R., Slettebak A., Wagner, R.M., 1998, A\&A, 339, 782
\bibitem{} Martin J.C., 2006, AJ, 131, 3047
\bibitem{} McMillan S.L.W., Hut P., 1996, ApJ, 467, 348
\bibitem{} Mikkola S., 1983, MNRAS, 203, 1107
\bibitem{} Murray S.D., Lin D.N.C., 1996, ApJ, 467, 728
\bibitem{} Oey M.S., Clarke C.J., 2005, ApJ, 620, L43
\bibitem{} Perets H.B., 2008, preprint (astro-ph/0802.1004)
\bibitem{} Perets H.B., 2009, ApJ, 690, 795
\bibitem{} Portegies Zwart S.F., 2000, ApJ, 544, 437
\bibitem{} Portegies Zwart S.F., McMillan S.L.W., 2002, ApJ, 576, 899
\bibitem{} Portegies Zwart S.F., Gaburov E., Chen H.-C., G\"urkan M.A., 2007, MNRAS, 378, L29
\bibitem{} Portegies Zwart S.F., Makino J.,  McMillan S.L.W., Hut P., 1999, A\&A, 348, 117
\bibitem{} Portegies Zwart S.F., McMillan S.L.W., Hut P., Makino J. 2001, MNRAS, 321, 199
\bibitem{} Portegies Zwart S.F., Baumgardt H., Hut P., Makino J., McMillan S.L.W., 2004, Nat, 428, 724
\bibitem{} Poveda A., Ruiz J., Allen C., 1967, Bol. Obs. Tonantzintla Tacubaya, 4, 86
\bibitem{} Przybilla N., Nieva M.F., Heber U., Butler K. 2008b, ApJ, 684, L103
\bibitem{} Przybilla N., Nieva M.F., Heber U., Firnstein M., Butler K., Napiwotzki R., Edelmann H., 2008a, A\&A, 480, L37
\bibitem{} Ramspeck M., Heber U., Moehler S., 2001, A\&A, 378, 907
\bibitem{} Rutledge R.E., Fox D.B., Shevchuk A.H., 2008, ApJ, 672, 1137
\bibitem{} Santos J.F.C., Bonatto C., Bica E., 2005, A\&A, 442, 201
\bibitem{} Sesana A., Haardt F., Madau P., 2006, ApJ, 651, 392
\bibitem{} Sesana A., Haardt F., Madau P., 2007, MNRAS, 379, L45
\bibitem{} Schilbach E., R\"{o}ser S., 2008, A\&A, 489, 105
\bibitem{} Stone R.C., 1991, AJ, 102, 333
\bibitem{} Suzuki T.K., Nakato N., Baumgardt H., Ibukiyama A., Makino J., Ebisuzaki T., 2007, ApJ, 668, 435
\bibitem{} Tenjes P., Einasto J., Maitzen H.M., Zinnecker H., 2001, A\&A, 369, 530
\bibitem{} Tillich A. et al., 2009, preprint (astro-ph/0901.1030)
\bibitem{} Tobin W., Kaufmann J.P., 1984, MNRAS, 207, 369
\bibitem{} T\"{u}llmann R., Rosa M.R., Elwert T., Bomans D.J., Ferguson A.M.N., Dettmar
R.-J., 2003, A\&A, 412, 69
\bibitem{} Weidner C., Kroupa P., 2004, MNRAS, 348, 187
\bibitem{} Weidner C., Kroupa P., 2006, MNRAS, 365, 1333
\bibitem{} Wolff S.C., Strom S.E., Dror D., Venn K., 2007, AJ, 133, 1092
\bibitem{} Yu Q., Tremaine S., 2003, ApJ, 599, 1129
\bibitem{} Yungelson L.R., van den Heuvel E.P.J., Vink J.S., Portegies Zwart S.F., de Koter A., 2008, A\&A, 477, 223
\bibitem{} Zinnecker H., Yorke H.W., 2007, ARA\&A, 45, 481

\end{thebibliography}
\end{document}